\documentclass[multphys,vecphys]{svmult}

\usepackage{makeidx}         
\usepackage{graphicx}        
\usepackage{multicol}        
\usepackage[bottom]{footmisc}

\makeindex             

\def\la{\mathrel{\hbox to 0pt{\lower 3.5pt\hbox{$\mathchar"218$}\hss}
      \raise 1.5pt\hbox{$\mathchar"13C$}}}
\def\ga{\mathrel{\hbox to 0pt{\lower 3.5pt\hbox{$\mathchar"218$}\hss}
      \raise 1.5pt\hbox{$\mathchar"13E$}}}

\begin{document}

\title*{Techniques for compact source extraction on CMB maps}
\author{R. B. Barreiro\inst{1}}
\institute{Instituto de F\'\i sica de Cantabria, CSIC - Universidad de
Cantabria, Avda. de los Castros s/n,\\
39005 Santander, Spain \\
\texttt{barreiro@ifca.unican.es}}
%
%
\maketitle

The detection of compact sources embedded in a background is a very
common problem in many fields of Astronomy. In these lecture notes we
present a review of different techniques developed for the detection
and extraction of compact sources, with a especial focus on their
application to the field of the cosmic microwave background
radiation. In particular, we will consider the detection of
extragalactic point sources and the thermal and kinematic
Sunyaev-Zeldovich effects from clusters of galaxies.

\section{Introduction}
\label{sec:introduction}
\begin{figure}
\centering
\includegraphics[angle=270,width=8cm]{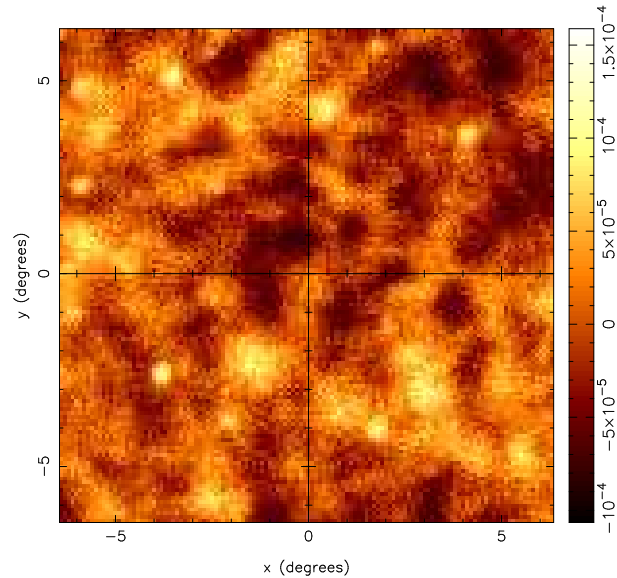}
\caption{Simulation of the 44 GHz Planck frequency channel in a small
patch of the sky, containing CMB, Galactic foregrounds, extragalactic
point sources and instrumental noise. The point sources can be seen as
localised objects embedded in the background. The Planck
Mission~\cite{bb:planck} is a satellite of the European Space Agency
to be launched in 2007 that will provide with multifrequency
observations of the whole sky with unprecedented resolution and
sensitivity.}
\label{fig:example_planck}
\end{figure}

Observations of an astrophysical signal in the sky are usually
corrupted by some level of contamination (called {\it noise} or {\it
background}), due to other astrophysical emissions and/or to the
detector itself. A common situation is that the signals of interest
are spatially well-localised, i.e. each of them covers only a small
fraction of the image, but we do not know a priori its position and/or
its amplitude. Some examples are the detection of extragalactic
sources in cosmic microwave background (CMB) observations (see
Fig.~\ref{fig:example_planck}), the identification of local features
(emission or absorption lines) in noisy one-dimensional spectra or the
detection of objects in X-ray images. It is clear that our ability to
extract all the useful information from the image will critically depend
on our capacity to disentangle the signal(s) of interest from the
background.

\begin{figure}
\centering
\includegraphics[height=\hsize]{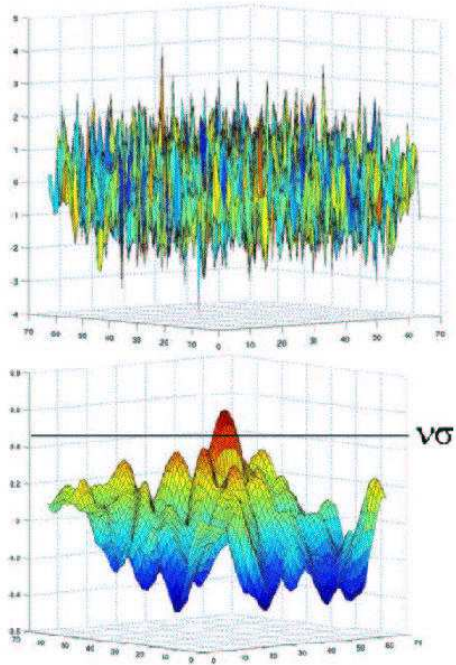}
\caption{This example illustrates the importance of filtering. A
source with a Gaussian profile has been placed in the centre of the
image in a background of white noise. The source can not
be distinguished in the original image (left panel), however, after filtering
(right panel), the source is enhanced over the background fluctuations.}
\label{fig:white_noise}
\end{figure}

The process to detect a localised signal in a given image
usually involves three different steps, which are not
necessarily independent:

1.- Processing: some processing of the data (commonly linear filtering)
is usually performed in order to amplify the searched signal over the
background. This is an important step because in many cases the
signals are relatively weak with respect to the background and 
it becomes very difficult to detect them in the original image.
This is illustrated in Figure~\ref{fig:white_noise}: the top panel
shows a simulation of white noise where a source with a Gaussian
profile has been added in the centre of the map; the bottom panel gives the
same simulation after filtering with the so-called matched filter. It
becomes apparent that the source was hidden in the original image
whereas it has been enhanced in the filtered image.

2.- Detection: we need a detection criterion, the {\it detector}, to decide
if some structure in the image is actually a real signal or if it is due to
the background. A very simple and widely used detector in Astronomy is
{\it thresholding}: if the intensity of the image is above a given value
(e.g. 5$\sigma$, where $\sigma$ is the dispersion of the map), a
detection of the signal is accepted, otherwise one assumes that
only background is present. In the example of
Fig.~\ref{fig:white_noise}, we see that several peaks appear in the
filtered image (right panel) but only one of them is above the
considered threshold $\nu \sigma$. Therefore, in this case, we would
accept only the highest peak as a true signal. 
Note that thresholding uses only the intensity of the data to make the
decision, however other useful information could also be included in
order to improve the detector (e.g. curvature, size, etc.).

3.- Estimation: a procedure must be established to estimate the
parameters (amplitude, size, position...) characterising the detected
signal. For instance, a simple possibility is to estimate the required
parameters by fitting the signal to its theoretical profile. 

The aim of these lecture notes is to present the problem of the
extraction of localised signals (compact sources) in the context of
CMB Astronomy and to review some of the methods developed to deal with
it. In section~\ref{sec:microwave_sky}, we outline the problem of
component separation in CMB observations. Section~\ref{sec:sources}
reviews some of the techniques developed for extraction of point
sources, including, among others, the matched filter and the Mexican
Hat Wavelet. Sections~\ref{sec:tsz} and~\ref{sec:ksz} deal
with the extraction of the thermal and kinematic Sunyaev-Zeldovich
effects in multifrequency microwave observations,
respectively. Section~\ref{sec:statistical} briefly discusses some
techniques for the extraction of statistical information from
undetected sources. Finally, in section~\ref{sec:conclusions} we
present our conclusions.

\section{The microwave sky and the problem of component separation} 
\label{sec:microwave_sky}
Microwave observations contain not only the cosmological signal but
also Galactic foregrounds, thermal and kinetic Sunyev-Zeldovich (SZ)
effects from clusters of galaxies and emission from extragalactic
point sources \cite{bb:bou99,bb:teg99,bb:ban03,bb:ben03}. In addition,
they also contain some level of noise coming from the detector itself.
In order to recover all the wealth of information encoded in the CMB
anisotropies, it is crucial to separate the cosmological signal from
the rest of the components of the sky.  Moreover, the foregrounds
themselves contain very valuable information about astrophysical
phenomena~\cite{bb:dez99}. Therefore, the development of tools to
reconstruct the different components of the microwave sky is of great
interest, not only to clean the CMB signal but also to recover all the
useful information present in the foregrounds.

The main Galactic foregrounds are the synchrotron, free-free and
thermal dust emissions. The synchrotron emission is due to
relativistic electrons accelerated in the Galactic magnetic field. The
free-free emission is the thermal bremsstrahlung from hot electrons
when accelerated by ions in the interstellar gas. The observed dust
emission is the sum over the emission from each dust grain along the
line of sight (dust grains in our galaxy are heated by the
interstellar radiation field, absorbing UV and optical photons and
re-emitting the energy in the far infrared).  In addition, there is
some controversial about the presence of an anomalous foreground at
microwave
frequencies~\cite{bb:deo02,bb:muk02,bb:lag03,bb:fin04,bb:deo04} that
could be due to the emission of spinning dust
grains~\cite{bb:dra98,bb:dra98b}.

The thermal SZ effect~\cite{bb:sz70,bb:sz72} is a spectral 
distortion of the blackbody spectrum of the CMB produced
by inverse Compton scattering of microwave photons by hot electrons in
the intracluster gas of a cluster of galaxies. In
addition, the radial peculiar velocities of clusters also produce secondary
anisotropies in the CMB via the Doppler effect, known as the
kinetic SZ effect~\cite{bb:sz80}. For a review on
the SZ effect, see~\cite{bb:bir99}.

\begin{figure}
\centering
\includegraphics[angle=270,width=8cm]{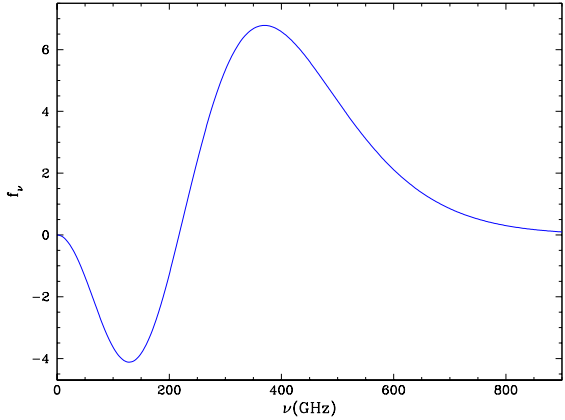}
\caption{Frequency dependence of the thermal SZ effect (in arbitrary
units)}
\label{fig:sz}
\end{figure}

The thermal SZ effect has a distinct spectral signature. It produces a
temperature decrement below 217 GHz and an increment above that
frequency. The change in intensity (see Fig.~\ref{fig:sz}) is given by
\begin{equation}
\Delta I = \frac{2(kT_o)^3}{(hc)^2}\frac{x^4e^x}{(e^x-1)^2} y_c \left[ 
x\coth \frac{x}{2}-4\right] \quad , \quad x=\frac{h\nu}{kT_o}
\end{equation}
where 
$y_c \equiv \frac{k{\sigma}_T}{m_e}\int dl\ T_e n_e$
is the Compton parameter and is a function
of the electron density $n_e$ and temperature $T_e$.
This distinct frequency dependence can be used in multifrequency
observations to separate the thermal SZ effect from the rest of
components of the microwave sky and, in particular, from the CMB.

The Doppler shift induced by the kinetic SZ effect in the CMB 
temperature fluctuations is given by: 
\begin{equation}
\frac{\Delta T}{T}=-\tau\frac{v_r}{c}
\end{equation}
where $v_r$ is the radial velocity of the cluster and
$\tau={\sigma_T}\int{n_e}dl$ is the optical depth.

The thermal and kinetic SZ effects imprint anisotropies in the CMB at
scales below a few arcminutes. Therefore, they are compact sources
whose shape is given by 
the convolution of the beam of the experiment with the cluster
profile. It is anticipated that will be very difficult to detect the
kinetic SZ effect and 
to separate it from the cosmological signal. This is due to the fact
that it has the same frequency dependence as the CMB (since it is just
a Doppler shift). Moreover, it is a very weak effect, around one order of
magnitude lower than the thermal effect.
The SZ effect is a very useful cosmological probe. Future SZ surveys
will allow one to obtain very valuable information about some of the
cosmological parameters, such as $H_0$, $\Omega_m$,
$\Omega_\Lambda$ and $\sigma_8$ (for a review see~\cite{bb:car02})

Emission from extragalactic point sources is an important contaminant
for high-resolution CMB experiments. By \emph{point source} is meant
that the typical angular size of these objects is much smaller than
the resolution of the experiment (which is usually the case in CMB
observations) and therefore they appear in the data as point-like
objects convolved with the beam of the instrument. There are two main
source populations: radio sources, which dominate at lower frequencies
($\la$ 300 GHz) and far-IR sources which give the main contribution at
higher frequencies ($\ga$ 300 GHz).  These populations consist mainly
of compact AGN, blazars and radio loud QSOs in the radio and of
inactive spirals galaxies in the far-IR.  Different models for the
radio ~\cite{bb:tof98,bb:tuc04,bb:dez05,bb:gon05} and
infrared~\cite{bb:gui99,bb:dol03,bb:gra04} point source populations
have been proposed.  However, there are still many uncertainties with
regard to the number of counts and the spectral behaviour of these
objects, due to a lack of data at the frequency range explored by CMB
experiments. Therefore experiments such as Planck will provide with
unique information to understand the astrophysical processes taking
place in these populations of sources. An additional problem is the
heterogeneous nature of extragalactic point sources, since, among other
complications, each source has its own frequency dependence. Therefore
they can not be treated as a single foreground to be separated from
the other components by means of multifrequency observations.

There are basically two different approaches to perform component
separation. The first one tries to reconstruct simultaneously all the
components of the microwave sky whereas the second one focuses in just
one single component. The first type of methods include the Wiener
filter~\cite{bb:bou96,bb:teg96}, maximum-entropy
method~\cite{bb:hob98,bb:hob99,bb:sto02,bb:ben03,bb:bar04,bb:sto05}
and blind source
separation~\cite{bb:bac00,bb:mai02,bb:mai03,bb:del03,bb:bed05,bb:pat05}.
These methods usually assume that the components to be reconstructed
can be factorised in a spatial template times a frequency dependence
(but see~\cite{bb:eri05} for a recent work where this assumption is
not necessary). This assumption is correct for the CMB and the SZ
effects but it is only an approximation for the Galactic
foregrounds. In addition, point sources can not be factorised in this
way and therefore these techniques are not well-suited for extracting
this contaminant. Regarding the second approach, it consists on
methods designed to extract a particular component of the sky. For
instance, the blind EM algorithm of ~\cite{bb:mar03} or the internal linear
combination of ~\cite{bb:ben03,bb:teg03,bb:eri04} try to recover only the CMB
component. Moreover, this type of approach is especially useful for the
detection of localised objects such as extragalactic point sources or
the SZ effects. In these lectures we will describe some of these
methods, that have been developed with the aim of extracting compact
sources from microwave observations.

\section{Techniques for extraction of point sources}
\label{sec:sources}

The most common approach to detect point sources embedded in a
background is probably linear filtering. A linearly filtered image
$w(x)$ is obtained as the convolution\footnote{Strictly speaking, the
filtered image can be written as a convolution provided the filter is
linear and spatially homogeneous (see e.g.~\cite{bb:her02d})} of the data
$y(x)$ with the filter $\psi(x)$:
\begin{equation}
w(x)=\int{y(u)\psi(x-u)du}
\end{equation}
Note that those parts of the data that resemble the shape of the
filter will be enhanced in the filtered map. Therefore, the filter
should have a similar profile to that of the sought signal.
Equivalently, we can work in Fourier space:
\begin{equation}
w(x)=\int{y(q)\psi(q)e^{-iqx}dq}
\end{equation}
where $f(q)$ denotes the Fourier transform of $f$. From the previous
equation, we see that the filter favours certain Fourier modes of the
data. 

In principle, it is equivalent to perform the filtering in real or
Fourier space. However, from the practical point of view, direct
convolution is a very CPU-time consuming operation. Therefore, working
in Fourier space, where a simple product is performed, is preferred.

Different linear filters have been proposed in the literature to
detect point sources in CMB maps, including the matched
filter~\cite{bb:teg98}, the Mexican hat
wavelet~\cite{bb:cay00,bb:vie01a,bb:vie03} , the scale adaptive
filter~\cite{bb:san01,bb:her02b}, the biparametric scale adaptive
filter~\cite{bb:lop05b} or the adaptive top hat
filter~\cite{bb:chi02}. In addition, non-linear techniques have also
been proposed, such as the Bayesian method of~\cite{bb:hob03} or the
non-linear fusion of~\cite{bb:lop05c,bb:lop05d}. In the next
subsections we give an overview of some of these techniques, including
applications to CMB simulated data.

\subsection{The matched filter}
Let us assume that we have a signal of amplitude $A$ at position $x_0$
embedded in a background of dispersion $\sigma$. The amplification
$\mathcal{A}$ of the signal obtained with a filter is given by:
%
\begin{equation}
\mathcal{A}=\frac{w(x_0)/\sigma_w}{A/\sigma}
\end{equation}
where $w(x_0)$ is the value of the filtered map at the position of the
source and $\sigma_w$ is the dispersion of the filtered map.
Therefore, if the amplification is greater than one, the contrast
between the signal and the background has been increased in the
filtered map, improving the chances of detecting the source with
respect to the original data. This is the main idea behind filtering:
it puts you in a better position to detect the sources.

The {\it matched filter} (MF) is defined as the linear filter that
gives maximum amplification of the signal. As an example, we will
outline how to construct the MF for a source $s(x)$ with spherical
symmetry (a more detailed derivation can be found
e.g. in~\cite{bb:her02d}). Let us consider a set of 2-dimensional data
$y(\vec{x})$:
\begin{eqnarray}
y(\vec{x})&=&s(x)+n(\vec{x})
\nonumber \\
s(x)&=&A\tau(x)
\end{eqnarray}
where $\vec{x}$ is a 2-dimensional vector of position
and $x=|\vec{x}|$. The source is characterised by a (spherically
symmetric) profile $\tau(x)$ and an amplitude $A=s(0)$. $n(\vec{x})$ is
the noise (or background) contribution which, for simplicity, is
assumed to be a homogeneous and isotropic random field with zero mean and
characterised by a power spectrum $P(q)$ ($q=|\vec{q}|$), i.e.,
\begin{equation}
\langle n(\vec{q}) n^*(\vec{q'}) \rangle = P(q) \delta^2 (\vec{q}-\vec{q'})
\end{equation}
where $n(\vec{q})$ is the 2-dimensional Fourier transform.

Let us introduce a filter $\psi$ with spherical symmetry. The filtered
field $w$ is given by:
\begin{equation}
w(\vec{x}) = \int{y(\vec{q})\psi(q)e^{-i\vec{q}\vec{x}}d\vec{q}}
\end{equation}

It can be shown that the filtered field at the position of the source
(for simplicity we will assume that the source is at the origin) is
given by:
\begin{eqnarray}
w(\vec{0})&=& 2 \pi \int q s(q) \psi(q) dq, \\
\nonumber
\end{eqnarray}
whereas the variance of the filtered field is obtained as:
\begin{equation}
\sigma_w^2 = 2 \pi \int q P(q) \psi^2(q) dq \\
\end{equation}
We want to find the filter that satisfies the following two
conditions: 
\begin{enumerate}
\item $\langle w(\vec{0})\rangle = A $
\item $\sigma_w^2$ is a minimum with respect to the filter $\psi$
\end{enumerate}
The first condition means that the filter is an unbiased estimator of
the amplitude of the source and gives straightforwardly the
constraint: 
\begin{equation}
\int q\tau(q)\psi(q)dq = \frac{1}{2 \pi}
\end{equation}
In order to minimise the variance of the filtered map (condition 2)
including the previous constraint, we introduce a Lagrange
multiplier $\lambda$:
\begin{equation}
\mathcal{L}(\psi)=\sigma_w^2(\psi)+\lambda \left[\int q \tau(q)
\psi(q) dq -\frac{1}{2 \pi} \right]
\end{equation}
Taking variations with respect to $\psi$ and setting the result equal
to zero, we find the matched filter:
\begin{eqnarray}
\psi(q)&=&k\frac{\tau(q)}{P(q)} \\
k&=&\left[2 \pi \int q \frac{\tau^2(q)}{P(q)} dq \right]^{-1}
\end{eqnarray}
Note that the matched filter is favouring those modes where the
contribution of the signal ($\tau$) is large and that of the noise
($P$) is small.

Assuming simple models for the Galactic foregrounds,~\cite{bb:teg98} has
given an estimation of the catalogue of point sources that Planck will
produce. According to this work, the number of sources detected by
Planck above a 5$\sigma$ level in a sky area of 8 sr will range from
around 650 for the 70 GHz channel to around 38000 at 857~GHz.

The matched filter has also been applied for the detection of point
sources in the 1st-year WMAP data~\cite{bb:ben03}. Using this
technique, a catalogue of 208 extragalactic point sources has been
provided.

\subsection{The Mexican Hat Wavelet}

Wavelet techniques are very versatile tools that only recently have
been applied to the analysis of CMB maps. The main property that makes
wavelet transforms so useful is that they retain simultaneously
information about the scale and position of the image. This means that
we can study the structure of an image at different scales without
loosing all the spatial information (as it occurs in the case of the
Fourier transform). There is not a unique way to construct a wavelet
transform (see e.g.~\cite{bb:dau92,bb:bur98}). In this section we will
focus on one particular wavelet, the Mexican Hat Wavelet (MHW), that
has been successfully implemented for the detection of point sources
with a Gaussian profile in CMB simulated
observations~\cite{bb:cay00,bb:vie01a,bb:vie03}.
\begin{figure}
\centering
\includegraphics[angle=0,width=8cm]{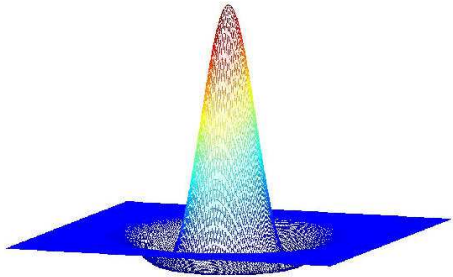}
\caption{The Mexican Hat Wavelet in 2 dimensions. }
\label{fig:mhw}
\end{figure}
The MHW is the second derivative of the Gaussian function (see
Fig.~\ref{fig:mhw}): 
\begin{equation}
\psi(x) = \frac{1}{\sqrt{2\pi}}\left[ 2 - \left( \frac{x}{R} \right)^2
\right] \exp\left(-\frac{x^2}{2R^2}\right)
\end{equation}
and in Fourier space is given by:
\begin{equation}
\psi(q) \propto (qR)^2 \exp\left(-\frac{(qR)^2}{2}\right)
\end{equation}
where $R$ is the scale of the MHW, a parameter that determines the
width of the wavelet. Note that $\psi(x)$ -- and wavelet functions in
general -- are compensated, i.e., the integral below the curve is
zero.  When filtering the data with the MHW, this property helps to
remove contributions of the background with a scale of variation
larger than the one of the wavelet.

\begin{figure}
\centering
\includegraphics[angle=270,width=\hsize]{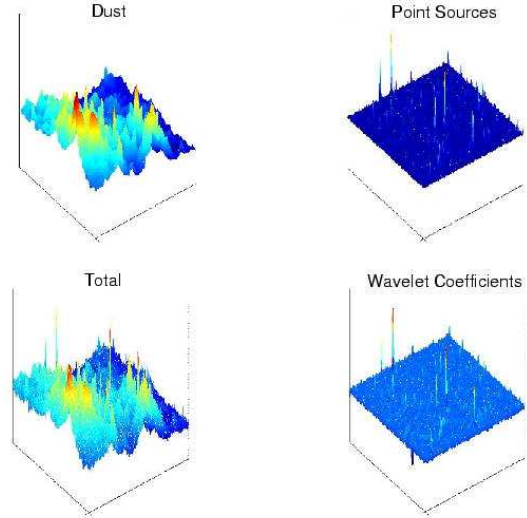}
\caption{An example of the performance of the MHW for a simulation of
the Planck 857 GHz channel (see text for details).}
\label{fig:canal857}
\end{figure}

The method to detect point sources is based on the study of the
wavelet coefficients map (i.e. the image convolved with the MHW) at a
given scale. Those wavelet coefficients above a fixed threshold are
identified as point source candidates. The reason why this works well, it
is because point sources are amplified in wavelet space. This can be
easily seen in Fig.~\ref{fig:canal857}, which shows a graphical
example of the performance of the MHW for a simulation of the Planck
857 GHz channel. 
Three of the panels correspond to the dust emission (top-left), which
is the dominant contaminant at this frequency, the emission of the
extragalactic point sources (top-right) and the total emission at this
frequency (bottom-left) including the Galactic foregrounds, CMB, SZ
effect, extragalactic point sources and instrumental noise. The last
panel (bottom-right) shows the total emission map after convolution
with a MHW at a certain scale $R_0$. It is clear that in the wavelet
coefficients map a large fraction of the background has been removed
and that the signal of the point sources has been enhanced. Therefore,
the detection level in wavelet space is greater than the detection
level in real space:
\begin{equation}
\frac{\omega(R)}{\sigma_\omega(R)} > \frac{A}{\sigma}
\end{equation}
where $\omega(R)$ is the wavelet coefficient at scale $R$ at the
position of the source, $\sigma_\omega$ is the dispersion of the
wavelet coefficients map, $A$ is the amplitude of the source and
$\sigma$ is the dispersion of the real map.

\begin{figure}
\centering
\includegraphics[angle=270,width=\hsize]{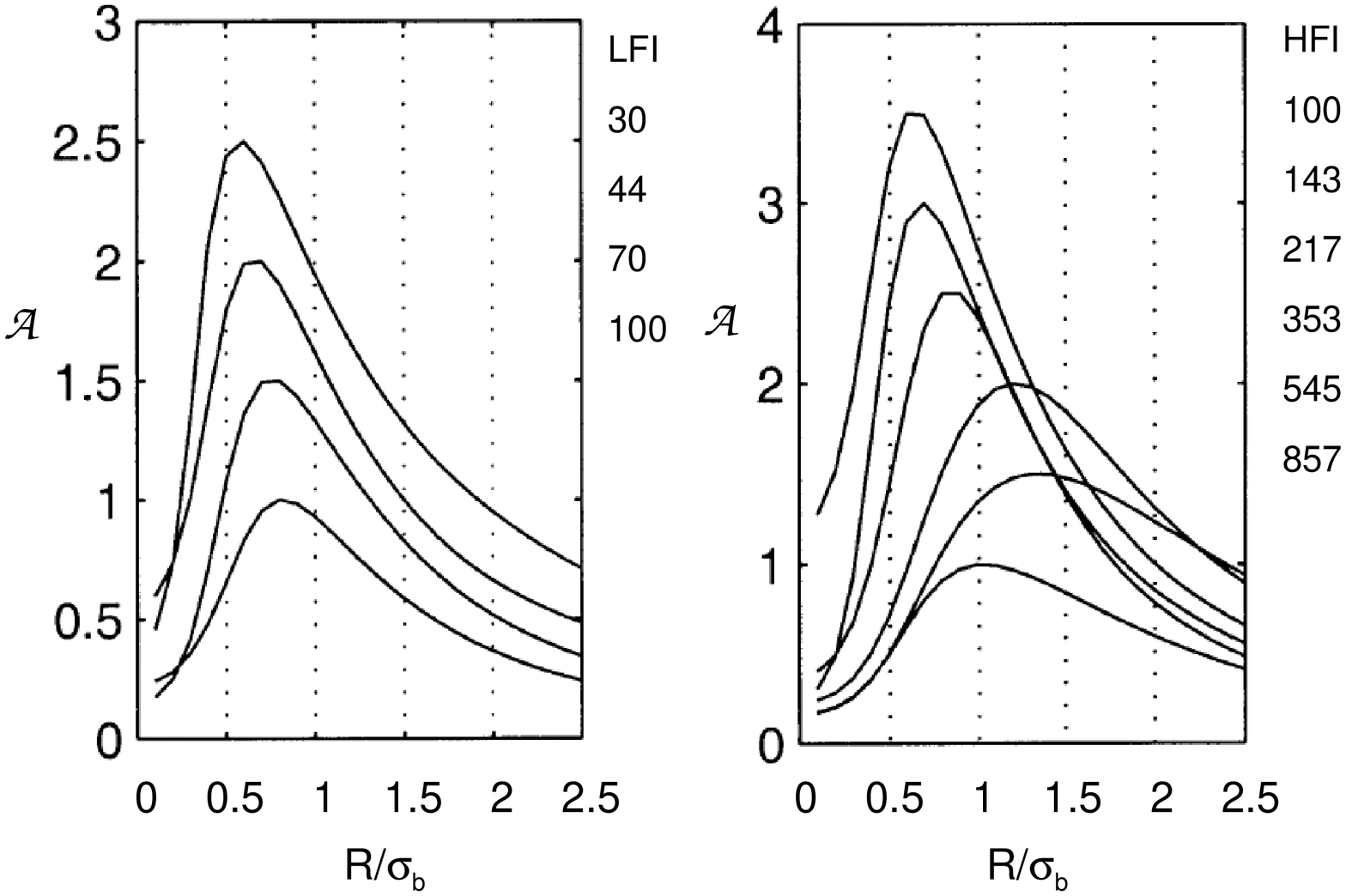}
\caption{Amplification (in arbitrary units) versus wavelet scale R (in
units of the beam dispersion) for a typical region of the sky (the 100
GHz channel of the Low Frequency Instrument has been withdrawn from
the final Planck payload due to financial shortage).}
\label{fig:optimal_scale}
\end{figure}

Note that the amplification (defined as the ratio between the
detection level in wavelet space and the detection level in real
space) depends on the wavelet scale $R$. In fact, for a given image,
there exists an optimal scale $R_0$ that gives maximum amplification
for the point sources and that can be determined from the data. For a
point source convolved with a Gaussian beam of dispersion $\sigma_b$,
the value of the wavelet coefficient $\omega(R)$ at the position of
the source is given by
\begin{equation}
\omega(R) = 2 R \sqrt{2 \pi} A
\frac{(R/\sigma_b)^2}{\left[1+(R/\sigma_b)^2\right]^2}  
\label{eq:w(R)}
\end{equation}
whereas the dispersion of the wavelet coefficients map at scale $R$ is 
\begin{equation}
\sigma^2_\omega(R) = 2 \pi R^2 \int{P(q)|\psi(qR)|^2 q dq}
\end{equation}
where $P(q)$ is the power spectrum of the background. Taking the
previous expressions into account, one can obtain the optimal scale
$R_0$ by maximising the amplification $\mathcal{A}$ versus
$R$. Fig.\ref{fig:optimal_scale} shows the amplification of the signal
versus the scale for simulated observations of the Planck Low and High
Frequency Instruments (LFI and HFI). Note that the optimal scale is close to
$\sigma_b$. This is expected, since this is the scale that
characterises the source, but the value of $R_0$ will also depend on
the background. For instance, if the noise contribution is more
important at scales smaller than the one of the source, this will tend
to move the optimal scale to values greater than $\sigma_b$ and
vice versa. Although the MHW will produce, in general, slightly less
amplification than the MF, it has the advantage of being an analytical
function, which greatly simplifies the use of this technique.

The procedure to detect point sources is as follows. First, the
optimal scale $R_0$ is obtained and the data are filtered with a MHW
of scale $R_0$. Those pixels $k$ above a given threshold are
identified as point sources.  The amplitude of the detected sources is
then estimated using a multiscale $\chi^2$ fit for each pixel $k$:
\begin{equation}
\chi^2_k = \sum_{i,j} (\omega_{i,k}^{th} - \omega_{i,k}^o)^t
V_{ij}^{-1}(\omega_{j,k}^{th} - \omega_{j,k}^o) 
\end{equation}
where $V$ is the covariance matrix between the different scales $i,j$
and $\omega_{i,k}^{th}$, $\omega_{i,k}^o$ correspond, respectively, to
the theoretical (given by equation ~\ref{eq:w(R)}) and observed
wavelet coefficient at scale $i$ and position $k$. Four scales are
used for the $\chi^2$. Note that this fit can also be used to discard
point source candidates: if a source candidate at pixel $k$ does
not have an acceptably low value of $\chi^2$ could be rejected as a point
source. 

The MHW technique has been implemented to deal with simulated Planck
observations in flat patches of the sky~\cite{bb:vie01a}
and also on the whole sphere~\cite{bb:vie03}. The
procedure to detect point sources on spherical data is very similar to
the one outlined before, but, in this case, the Spherical Mexican Hat
Wavelet is used to convolve the data, which is given by~\cite{bb:mar02}
\begin{eqnarray}
\label{eqSMHW}
   \Psi_S(y,R) &=&
  \frac{1}{\sqrt{2\pi}N(R)}{\Big[1+{\big(\frac{y}{2}\big)}^2\Big]}^2
  \Big[2 - {\big(\frac{y}{R}\big)}^2\Big]e^{-{y}^2/2R^2},
\nonumber
\\
      N(R) & \equiv & R{\Big(1 + \frac{R^2}{2} + \frac{R^4}{4}\Big)}^{1/2}.
\end{eqnarray}
$y$ is the distance to the tangent plane which is given by $y\equiv
2\tan \frac{\theta}{2}$, where $\theta$ is the latitude angle.
Another important point when dealing with spherical data is the
estimation of the optimal scale. In CMB observations, the contaminants
can be very anisotropic in the sky. This means
that the properties of the background change significantly for
different areas of the sky and therefore the optimal scale for
filtering with the SMHW has to be obtained locally. This is simply
done by projecting small patches of the sky on a plane and obtaining
the optimal scale for each of them. The maps are then convolved with
SMHW of different scales and the detection is performed on the sphere
but with the corresponding $R_0$ for each patch.

Table~\ref{catalogue} shows the catalogue predicted by~\cite{bb:vie03}
using simulated Planck observations of the whole sky. In addition to
CMB and point sources, Galactic foregrounds, thermal SZ and
instrumental noise were included in the simulations. Using the
recovered catalogue, mean spectral indices can also be estimated with
good accuracy. The SMHW has also been adapted to deal with realistic
asymmetric beams as those expected for the Planck Mission.

\begin{table}
\centering
\caption{ Predicted point source catalogue using the SMHW from
simulated Planck observations. The point sources have been detected
outside a Galactic cut that varies from channel to channel (from no
cut at the lowest frequency channels up to a Galactic cut with $b=25$
degrees for 857 GHz). The different columns correspond to: Planck
frequency channel, number of detections (above the minimum flux),
minimum flux, mean error, mean bias, number of optimal scales needed
for the algorithm and completeness of the catalogue above the minimum
flux (see~\cite{bb:vie03} for details).}
\label{catalogue}
\begin{tabular}{ccccccc}
\hline\noalign{\smallskip}
 Frequency & ~Number~ & ~Min. Flux (Jy)~ & 
         ~$\bar{E}(\%)$~  & ~$\bar{b}(\%)$ & 
         ~$N_{R_o}$~ & ~Completeness~\\
	(GHz) & & & & & & (\%) \\
\noalign{\smallskip}\hline\noalign{\smallskip}
	 857      & 27257 & 0.48 & 17.7 &  -4.4 &  17 & 70 \\
	 545      &  5201 & 0.49 & 18.7 &   4.0 &  15 & 75 \\
	 353      &  4195 & 0.18 & 17.7 &   1.4 &  10 & 70 \\
	 217      &  2935 & 0.12 & 17.0 &  -2.5 &    4 & 80 \\
	 143      &  3444 & 0.13 & 17.5 &  -4.3 &    2 & 90 \\
	 100      &  3342 & 0.16 & 16.3 &  -7.0 &    4 & 85 \\
	  70      &  2172 & 0.24 & 17.1 &  -6.7 &    6 & 80 \\
	  44      &  1987 & 0.25 & 16.4 &  -6.4 &    9 & 85 \\
	  30      &  2907 & 0.21 & 18.7 &   1.2 &    7 & 85 \\
\noalign{\smallskip}\hline
\end{tabular}
\end{table}

It is also interesting to note that the MHW technique has been
combined with the maximum-entropy method~\cite{bb:vie01b}. Using
Planck simulated data it was shown that the joint method improved the
quality of both the reconstruction of the diffuse components and the
point source catalogue.

Finally, we would like to point out that the MHW has been used for the
detection of objects in X-ray images~\cite{bb:dam97a,bb:dam97b} and,
more recently, in SCUBA~\cite{bb:barn04,bb:knu05} and Boomerang
data~\cite{bb:cob03}. 

\subsection{The Neyman-Pearson detector and the biparametric scale
adaptive filter} 

As mentioned in section~\ref{sec:introduction}, filtering helps the
detection process because it amplifies the sought signal over the
background. However, whether we filter or not, we still need a
detection criterion -- the detector -- to decide if a given signal
belongs to the background or to a true source. In addition, the final
performance of the filter will clearly depend on the choice of the
detector. A criterion that has been extensively used in Astronomy is
thresholding, i.e., those pixels of the data above a given value
(e.g. 5$\sigma$) are identified as the signal. Thresholding has a
number of advantages, including its simplicity and the fact that it
has a precise meaning in the case of Gaussian backgrounds in the sense
of controlling the probability of spurious detections. However, it
only uses a limited part of the information contained in the data, the
intensity, to perform decisions.  

An example of detector -- based on the Neyman-Pearson rule -- that
takes into account additional information has been recently proposed
in~\cite{bb:bar03,bb:lop04,bb:lop05} for 1-dimensional signals
and~\cite{bb:lop05b} for the 2-dimensional case. The first step of the
procedure is to identify maxima as point source candidates. To decide
then whether the maxima are due to the presence of background on its
own or to a combination of background plus source, a Neyman-Pearson
detector is applied, which is given by (for 2-dimensional signals):
\begin{equation}
L(\nu, \kappa, \epsilon) \equiv
\frac{n(\nu,\kappa,\epsilon)}{n_b(\nu,\kappa,\epsilon)} \ge L_*
\end{equation}
where $n_b(\nu,\kappa,\epsilon) $d$\vec{x} $d$\nu $d$\kappa $d$\epsilon$ is
the expected number of maxima of the background in the intervals
$(\vec{x}, \vec{x}+$d$\vec{x})$, $(\nu, \nu+$d$\nu)$, $(\kappa,
\kappa+$d$\kappa)$ and $(\epsilon, \epsilon+$d$\epsilon)$, whereas
$n(\nu,\kappa,\epsilon)$d$\vec{x} $d$\nu $d$\kappa $d$\epsilon$ corresponds to
the same number in the presence of background plus source. $\nu$,
$\kappa$ and $\epsilon$ are the normalised intensity, normalised curvature and
normalised shear of the field respectively, and $\vec{x}$ is the spatial
variable.
For a homogeneous and isotropic Gaussian background and point sources
with spherical symmetry, it can be shown that the previous detector is
equivalent to~\cite{bb:lop05b}:
\begin{equation}
\varphi(\nu,\kappa) = a \nu + b \kappa  \ge \varphi_*
\end{equation}
where $a$ and $b$ are constants that depend on the properties of the
background and the profile of the source, and $\varphi_*$ is a constant
that needs to be fixed. Therefore, if the considered
maximum satisfies $\varphi \ge \varphi_*$, we decide that the signal is
present, otherwise we consider that the maximum is due to the presence
of only background. 

Using this detector,~\cite{bb:lop05b} compares the performance of
different filters. In order to do this, $\varphi_*$ is fixed to produce
the same number of spurious sources for all the filters and then the
number of true detections are compared. In their study they consider
the matched filter, the Mexican hat wavelet, the scale adaptive filter
and the biparametric scale adaptive filter (BSAF). In addition, the
scale of the filter is allowed to vary (similarly to what is done in
the Mexican hat wavelet technique) through the introduction of a
parameter $\alpha$.  For the case of a background with a scale-free
power spectrum ($P(q) \propto q^{-\gamma}$) and a source with Gaussian
profile, the BSAF is given by:
\begin{equation}
\psi_{BSAF} (q) = N(\alpha) z^\gamma e^{z^2/2}(1+cz^2), ~~~~ z\equiv q
\alpha R 
\end{equation}
where $c$ is a free parameter. $\alpha$ and $c$ are optimised in order
to produce the maximum number of true detections given a fixed number
of spurious detections. For $c=0$ and $\alpha=1$ the MF is
recovered. Using this approach the performance of the considered
filters has been studied for the case of a Gaussian white noise
background. The results predict that, in certain cases, the BSAF can
obtain up to around 40 per cent more detections than the other
filters. Although in CMB observations the background is not usually
dominated by white noise, this case can also be of interest to detect
point sources on CMB maps that have been previously processed using a
component separation technique such as the maximum-entropy
method~\cite{bb:vie01b}. In this case, the expected contribution of
foregrounds and CMB is subtracted from each of the frequency maps,
leaving basically the emission of extragalactic point sources and white
noise (as well as some residuals). In this type of maps, the
application of this technique could be useful. In any case, a test of
the BSAF on realistic CMB simulations would be necessary to establish
how well this approach would perform on real data.

\subsection{Bayesian approach to discrete object detection}

Many of the standard techniques for the detection of point sources are
based on the design of linear filters. However, other methods --
usually more complicated -- are also possible. For
instance,~\cite{bb:hob03} has recently proposed a Bayesian approach
for the detection of compact objects. The method is based on the
evaluation of the (unnormalised) posterior distribution
$\overline{Pr}(\theta|D)$ for the parameters $\theta$ that characterise the
unknown objects (such as position, amplitude or size), given the
observed data $D$. The unnormalised posterior probability is given in
terms of the likelihood Pr$(D|\theta)$ and the prior Pr($\theta$) as:
\begin{equation}
\overline{Pr}(\theta|D) \equiv Pr(D|\theta) Pr(\theta)
\end{equation}
Two different strategies are proposed for the detection of compact
sources: an exact approach that tries to detect all the objects
present in the data simultaneously and an iterative -- much faster --
approach (called McClean algorithm). In both cases an estimation of
the parameters of the sources as well as their errors are
provided. For both methods, a Markov-Chain Monte-Carlo technique is
used to explore the parameter space characterising the objects.

As an illustration of the performance of the method,~\cite{bb:hob03}
studies the performance of both algorithms for a simple example that
contains 8 discrete objects with a Gaussian profile embedded on a
Gaussian white noise background. The test image has 200$\times$200
pixels and the signal-to-noise ratio of the objects ranges from 0.25
to 0.5. Using the exact method, where the number of objects is an
additional parameter to be determined by the algorithm, all the
objects are detected with no spurious detections. However two of the
objects (which overlapped in the noiseless data) are identified as a
single detection. The results show that the parameters have been
estimated with reasonably good accuracy. Unfortunately, although this
method seems to perform very well, it is also very computationally
demanding, what can make it unfeasible in many realistic applications.

The iterative approach tries to detect the objects one-by-one. This
significantly reduces the CPU-time necessary for the algorithm whereas
provides a convenient approximation to the exact method. For the
simple example previously considered, the McClean algorithm provides
quite similar results to the exact approach, with only one less object
detected than the exact approach.

This approach is certainly very promising and can be very useful for
the detection of compact sources in future CMB data. However it
assumes the knowledge of the functional form of the likelihood, the
prior of the parameters and the profile of the objects, which will not
be known in many realistic situations. In addition, the presence of
anisotropic contaminants (such as instrumental noise or Galactic
foregrounds) would introduce additional complexity that would make the
algorithm more computationally demanding. \cite{bb:hob03} gives some
hints on how to deal with some of these problems. In any case, it
would be very useful to test the performance of the method under
realistic conditions in order to establish the real potentiality of
the technique.

\section{Techniques for extraction of the thermal SZ effect}
\label{sec:tsz}

Another important application of the compact source extraction
techniques is the detection of the thermal SZ effect due to galaxy
clusters in microwave observations. The resolution of most CMB
experiments (e.g. 5 arcminutes for the best Planck channels) is
usually not enough to resolve the structure of the clusters of
galaxies. Thus, the SZ emission appears in the CMB maps as compact
sources whose shape is given by the convolution of the beam of the
experiment with the profile of cluster. This means that most of the
techniques used for the detection of extragalactic point sources could
be easily adapted to detect the SZ effect (in one single map), just by
including the correct profile of the sought source in the
algorithm. This type of studies have been done for instance for the
SAF~\cite{bb:her02c}, the MF~\cite{bb:sch03} or the Bayesian
approach~\cite{bb:hob03} discussed in the previous section.

However, the thermal SZ effect has a very characteristic frequency
signature that can be used to extract this emission, provided that
multifrequency observations are available. One alternative to recover
the SZ emission is to apply a component separation technique -- such
as the maximum-entropy method, Wiener filter or blind source
separation -- that tries to recover simultaneously all the different
emissions of the microwave sky. The second possibility is to design
specific methods to extract the SZ signal that make use of the
multifrequency information
\cite{bb:her02a,bb:die02,bb:pie05}. Regarding this second strategy, we
will describe two of the methods that have been devised for the
extraction of the thermal SZ effect from multifrequency CMB
observations. Both methods have been tested using Planck simulated
data.

\subsection{Filtering techniques}

\cite{bb:her02a} presents different filtering techniques for the
detection of SZ clusters in multifrequency maps. Two alternative
strategies are proposed: a combination technique and the design of a
multifrequency filter (or multifilter). In both cases, the spatial
profile of the clusters are assumed to be known. In the combination
method, the individual frequency maps are linearly combined in an
optimal way. The weights of the linear combination can be determined
from the data and they are optimal in the sense of giving the maximum
amplification of objects that have the required spatial profile and
the correct frequency dependence. This combined map is then filtered
either with the MF or with the SAF constructed taking into account the
characteristics of this new map. In the second approach, each
frequency map is filtered separately but the filters are constructed
taken into account the cross-correlations between frequency channels
as well as the spectral dependence of the SZ effect. Then, the
filtered maps are added together. This second method can also be
implemented for two different kind of filters: the matched multifilter
and the scale-adaptive multifilter. \cite{bb:her02a} performs a
comparison of all these techniques finding that the matched
multifilter provides the best results. Another interesting point is
that the combination method is appreciably faster than the multifilter
technique, whereas it still detects a large fraction of the clusters
found by the matched multifilter.

Taking these results into account, we will discuss in more detail the
matched multifilter (MMF) approach\footnote{A similar technique has
also been independently developed by~\cite{bb:nas02}, but in the
context of the detection of extragalactic point sources in
multifrequency observations}. Let us consider a set of N observed maps
given by:
\begin{equation}
y_\nu (\mathbf{x}) = f_\nu A \tau_\nu(\mathbf{x}) + n_\nu
(\mathbf{x}), ~~~ \nu=1,...,N
\end{equation}
where, for illustration purposes, it is assumed that the SZ signal is
due to the presence of a single cluster located in the origin of the
image. The first term of the right hand side describes the
contribution of the sought signal (the thermal SZ effect in this case)
whereas $n_\nu$ is a generalised noise term that includes the sum of
all the other components present in the map. $f_\nu$ is the frequency
dependence of the thermal SZ effect (normalised to be 1 at a reference
frequency), $\tau_\nu$ is the shape of the cluster at each frequency
(i.e. the profile of the cluster convolved with the corresponding
antenna beam) and $A$ is the amplitude of the SZ effect at the
reference frequency.  For simplicity the profile of the cluster is
assumed to be spherically symmetric and is parameterised by a
characteristic scale -- the core radius $r_c$ -- but a generalisation
to more complex profiles can be easily done. The background is assumed
to be a homogeneous and isotropic random field with zero mean value
and cross-power spectrum $P_{\nu_1 \nu_2}(q)$ defined as
\begin{equation}
\langle n_{\nu_1}(\mathbf{q}) n^*_{\nu_2}(\mathbf{q}') \rangle
=P_{\nu_1 \nu_2}(q) \delta_D^2(\mathbf{q}- \mathbf{q}'), ~~~~ q \equiv
|\mathbf{q}| 
\end{equation}
where $n_\nu(\mathbf{q})$ is the Fourier transform of
$n_\nu(\mathbf{x})$ and $\delta_D^2$ is the 2-dimensional Dirac
distribution. 

The MMF is given (in matrix notation) by
\begin{equation}
\mathbf{\Upsilon}(q)=\alpha \mathbf{P}^{-1} F, ~~~ \alpha^{-1} = \int
d\mathbf{q} F^t \mathbf{P}^{-1}F
\end{equation}
where $F$ is the column vector $F=[f_{\nu} \tau_{\nu}]$ and
$\mathbf{P}^{-1}$ is the inverse of the cross-power spectrum matrix
$P\equiv[P_{\nu_1 \nu_2}]$.

The output map, where the detection of the SZ effect is finally
performed, is obtained by filtering each frequency map with its
corresponding filter $\psi_{\nu}$ and then adding together all the
filtered maps.  The detection is then performed by looking for regions
(5 or more pixels) of connected pixels above a 3$\sigma$
threshold. The maximum of the region determines the position of the
cluster. In addition, the MMF is constructed so that the value
of the intensity of the output map in the position of the source is an
unbiased estimator of the amplitude. Therefore, the estimated
amplitude of the SZ is simply given by the value of the output field
in the considered maximum. Another interesting point is that the scale
of the clusters $r_c$ will not be known a priori. To overcome this
problem, the data is multifiltered using different values of
$r_c$. When the scale of the cluster coincides with that of the
filter, the amplification of the signal will be maximum and that gives
an estimation of the core radius $r_c$.

\begin{figure}
\centering
\includegraphics[angle=0,width=6cm]{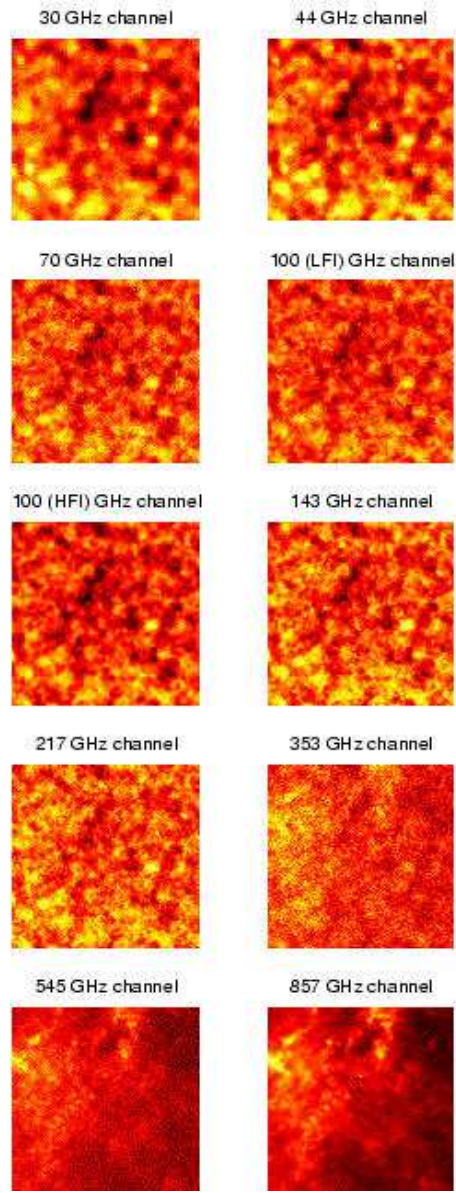}
\caption{Simulated Planck channels used to test the performance of the
MMF.}
\label{fig:data}
\end{figure}

The MMF has been tested on Planck simulated data of small patches
(12.8$^\circ$ $\times$ 12.8$^\circ$) of the sky containing CMB,
thermal and kinetic SZ effects, Galactic foregrounds (synchrotron,
free-free, thermal dust and spinning dust) extragalactic point sources
and instrumental noise. The simulated Planck data are shown in
Fig.~\ref{fig:data}. Note that the SZ emission of clusters is
completely masked by the rest of the components present in the data.

\begin{figure}
\centering
\includegraphics[angle=0,width=9.2cm]{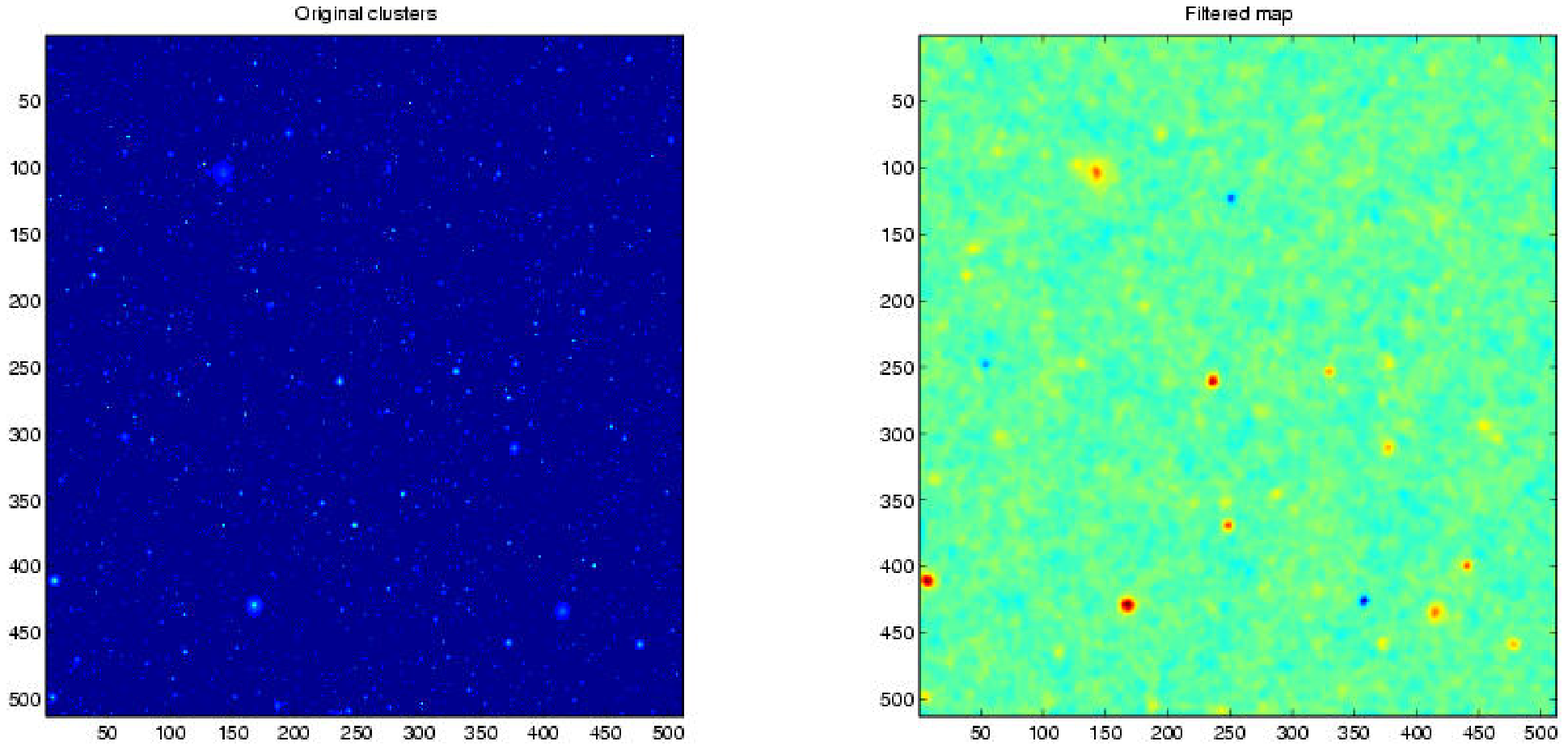}
\caption{The input (left panel) and reconstructed (right panel) SZ
effect after filtering the data with a MMF of $r_c=1$ pixel.} 
\label{fig:mmf_results}
\end{figure}

Fig.~\ref{fig:mmf_results} shows the input thermal SZ emission
included in the simulations and the reconstructed SZ map after
filtering the data with a MMF of $r_c$=1 pixel. Clusters with scales similar
to the chosen one are clearly visible in the output map.
\cite{bb:her02a} finds that the mean error in the determination of the
position of the clusters is around 1 pixel whereas the core radii are
determined with an error of 0.30 pixels. Regarding the determination
of the cluster amplitudes, the mean error is around 30 per cent for
the brightest clusters, whereas there is a bias in the estimation of
the weakest clusters. This bias can be understood since most weak
clusters will only reach the detection threshold if they are on top of
a positive fluctuation of the background, which will lead to an
overestimate of the amplitude. The bias could be reduced by improving
the method of amplitude estimation (for instance by performing a fit
to the profile of the cluster) that was simply given by the value of
the maximum of the detection. Using this technique it is expected that
Planck detects around 10000 clusters in 2/3 of the sky.

An extension of the multifilter technique to spherical data has been
carried out by~\cite{bb:sch04}. They test the
method on realistic simulations of Planck in the whole sky that, in
addition to the main microwave components, also include the effect of
non-uniform noise, sub-millimetric emission from celestial bodies of
the Solar system and Galactic CO-line radiation. It is again found
that the multifilter approach can significantly reduce the background,
allowing the cluster signal to be detected.

\subsection{Bayesian non-parametric technique}

\cite{bb:die02} proposes an alternative method to detect SZ
clusters in Planck data. The method has been also tested on the
simulated data set of Fig.~\ref{fig:data}. The procedure is as
follows. First of all, the frequency maps are significantly cleaned
from the most damaging contaminants. In particular, extragalactic
point sources are subtracted with the MHW and subsequently the
emission from dust and CMB is removed using the information of the 857
and 217 GHz channels respectively. The next step consists on obtaining
a map of the Compton parameter $y_c$ in Fourier space by maximising,
mode by mode, the posterior probability $P(y_c|d)$. Taking into
account Bayes' theorem, this probability is given by
\begin{equation}
P(y_c|d) \propto P(d|y_c)P(y_c) 
\end{equation}
In order to perform this maximisation we need to know the likelihood
function $P(d|y_c)$ and the prior $P(y_c)$. Since the residuals left in
the frequency maps are mainly dominated by the instrumental noise, the
likelihood can be well approximated by a multivariate Gaussian
distribution.  In addition, one needs to assume a form for the prior
$P(y_c)$. Using SZ simulations,~\cite{bb:die02} finds that the prior follows
approximately the form $P(y_c)\propto \exp(-|y_c|^2/P_{y_c})$ at each
k-mode, where $P{y_c}$ is the power spectrum of the SZ map. Taking
these results into account one gets, after maximising the posterior
probability, the following solution for the $y_c$ map at each mode:
\begin{equation}
y_c=\frac{\mathbf{dC}^{-1}\mathbf{R}^\dag}
{\mathbf{RC}^{-1}\mathbf{R}^\dag+P_{y_c}^{-1}} 
\end{equation}
where $d$ is the data, $R$ is the response vector (that includes the
information from the beam at each frequency and the frequency
dependence of thermal SZ effect) and $C$ is the cross-correlation
matrix of the residuals. This result coincides with the
multifrequency Wiener filter solution for the Compton $y_c$
parameter. Note that the recovered $y_c$ map will depend on the
assumed power spectrum $P_{y_c}$. However,~\cite{bb:die02} shows that the
final results do not depend significantly on the particular choice of
$P(y_c)$, provided its form satisfies some general conditions. We
would like to remark that this method does not need to make any
assumption about the profile of the SZ clusters.

\begin{figure}
\centering
\includegraphics[angle=0,width=9.2cm]{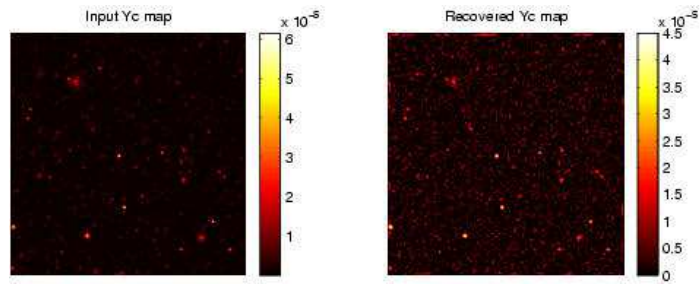}
\caption{The input (left panel) and reconstructed (right panel)
  Compton parameter map after applying the Bayesian non-parametric
  method.}
\label{fig:bayes_results}
\end{figure}

The recovered $y_c$ map after applying this approach to the Planck
simulated data of Fig.~\ref{fig:data} is shown in
Fig.~\ref{fig:bayes_results}. 
The detection and the estimation of the flux of clusters is performed
in the recovered map using the package SExtractor~\cite{bb:ber96},
which selects connected pixels above a given threshold. Using a
$3\sigma$ threshold, this method predicts that Planck will detect
around 9000 SZ clusters over 4/5 of the sky. In addition, it is shown
that the flux of the clusters is estimated with no significant bias,
which is important to carry out cosmological studies with the
recovered catalogue.

\section{Techniques for extraction of the kinetic SZ effect}
\label{sec:ksz}

The determination of peculiar velocities of individual clusters
through the kinetic SZ effect is a very challenging task. This is
mainly due to the facts that the kinetic SZ emission is very weak --
around 1 order of magnitude weaker than the thermal SZ effect --
and that it has the same frequency dependence as the CMB, meaning that
both signals can not be separated using only multifrequency
information. In addition, the presence of all the other components of
the microwave sky and the instrumental noise makes even more difficult
to detect this tiny signal. On the other side, we should take
advantage of some characteristics of the kinetic SZ that could help to
extract this emission. An important point is to use the available
information about the thermal SZ. Since both SZ effects are
produced by clusters of galaxies, there is a strong spatial
correlation between them. Multifrequency observations are also
important to separate the kinetic SZ signal from the components of the
microwave sky (except from the CMB). In particular, it is useful to
consider observations at the frequency of 217 GHz, where the
contribution of the thermal SZ is expected to be negligible. Finally,
the probability distribution of the kinetic SZ signal (expected to be
highly non-Gaussian) and its power spectrum are very different from
the ones of the cosmological signal, what could also be used to
separate it from the CMB fluctuations.

Due to the complexity of the problem, only a few methods have been
proposed and tested with the aim of extracting this emission from
microwave observations. For instance,~\cite{bb:hae96} studied the
performance of an optimal filter (which is actually a matched filter
for the cluster profile) to detect the kinetic SZ effect, concluding
that peculiar velocities could be measured only for a few fast moving
clusters at intermediate redshift. \cite{bb:hob03} also applied
their Bayesian algorithm to detect the kinetic SZ effect on CMB
simulations at 217 GHz, but including only CMB and instrumental noise
in the background. They claim that their technique is around twice as
sensitive as the optimal linear filter. An alternative approach has
been proposed by~\cite{bb:for04}, which makes use of spatial
correlation between the thermal and kinetic SZ effects. The method is
tested in ideal conditions, using as starting point a map of the
Compton parameter and a second map containing only CMB and kinetic SZ
emission. In these ideal conditions the method provides very promising
results. However, a detailed study of the performance of the method
under realistic conditions should be done before establishing the true
potentiality of this approach. Recently,~\cite{bb:her05} tested
a modification of the matched multifilter on Planck simulated data for
the detection of the kinetic SZ effect, that we discuss here in more
detail.

\subsection{The unbiased matched multifilter}

If multifrequency information is available, we can also construct a
MMF adapted to the kinetic SZ emission. As for the thermal SZ, the
shape of the sought source will be the convolution of the antenna beam
with the cluster profile but the frequency dependence will now follow
that of the kinetic SZ emission. However,~\cite{bb:her05} found
that the estimation of the kinetic SZ effect (and in fact that of the
thermal SZ effect) using the MMF is intrinsically biased. It can be
shown that this is due to the presence of two signals (the thermal and
kinematic SZ effects in this case) that have basically the same spatial
profile. Given the difference in amplitude of both effects, this
bias is negligible for the case of the thermal SZ but it can be very
important for the kinetic one. In order to correct this bias, a new
family of filters, the unbiased matched multifilter (UMMF) has been
constructed. For the case of the kinetic SZ effect, the UMMF is given
by~\cite{bb:her05}:
\begin{eqnarray}
\mathbf{\Phi} & =& \frac{1}{\alpha \gamma -
\beta^2}(-\beta\mathbf{F}+\alpha \tau) \nonumber \\
\beta &=& \int d\mathbf{q} \tau^t \mathbf{P}^{-1} \mathbf{F} , ~~~~~
\gamma = \int d\mathbf{q} \tau^t \mathbf{P}^{-1} \tau
\end{eqnarray}
These new multifilter leads to a slightly lower amplification of the
sources than the MMF, but is intrinsically unbiased. The UMMF has been
tested using Planck simulated data of small patches of the sky
including CMB, Galactic foregrounds (synchrotron, free-free, thermal
and spinning dust) and point sources. In order to test just the effect
of the intrinsic bias on the estimation of the amplitude, simplistic
simulations of clusters have been used, and the knowledge of the
profile and position of the clusters has been
assumed. 

\begin{figure}
\centering
\includegraphics[angle=0,width=\hsize]{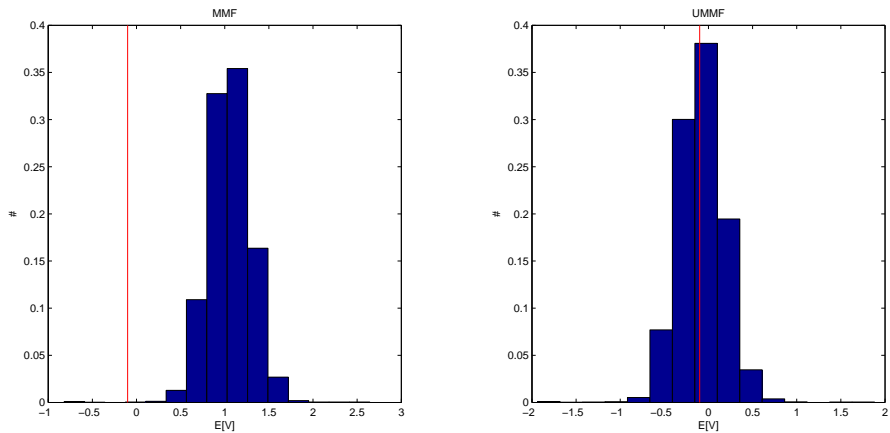}
\caption{Normalised histogram of the $V$ parameter using the MMF (left
panel) and the UMMF (right panel). The red vertical line indicates the
input value of $V$.}
\label{fig:ummf}
\end{figure}

Fig.~\ref{fig:ummf} shows the normalised histogram of the recovered
parameter $V=(v_r m_e c)/(k_B T_e)$ using the MMF and the UMMF
obtained from simulations that contained clusters with $r_c$=1.5
arcmin, $y_c=10^{-4}$ and $V=-0.1$. For a temperature of the electrons
of $T_e \simeq 5$keV, this value of $V$ corresponds to a radial
velocity along the line of sight of $v_r \simeq 300$ kms$^{-1}$. As
predicted, the estimation of the amplitude of the kinetic SZ effect is
strongly biased when using the MMF. However, this bias is corrected
when the data are filtered with the UMMF. \cite{bb:her05} also shows
that this result remains valid for smaller values of $y_c$ or $V$.
Unfortunately, the error in the determination of peculiar velocities
remains very large even for bright clusters. For instance, for
$y_c=10^{-4}$ and $T_e \simeq 5$keV the statistical error in the
determination of $v_r$ is $\sim$ 800 kms$^{-1}$. This means that
Planck will not be able to measure, in general, the peculiar
velocities of individual clusters, at least using just
UMMF. Nonetheless, since the UMMF provides an unbiased estimation of
$v_r$, it could be possible to measure mean peculiar velocities on
large scales by averaging over many clusters.

\section{Extraction of statistical information from undetected sources}
\label{sec:statistical}

In some cases the sought signal may be too weak to be individually
detected or we may have found the brightest sources but we are unable
to go down in flux. However, in this case, it may still be possible to
extract some valuable statistical information from the background of
sources. In this section we briefly discuss some works that have
pursued this objective.

The differential number counts of extragalactic point sources are
usually parameterised as 
\begin{equation}
n(S) = kS^{-\eta} ~, ~~~S > 0
\end{equation}
where $S$ is the flux of the source and $k$ and $\eta$ are the
normalisation and slope parameters respectively. \cite{bb:pie03} uses
the information of the high order moments of simulated data containing
CMB, noise and residual point sources, to estimate these two
parameters. \cite{bb:her04} determines $k$ and $\eta$ by fitting the
characteristic function of the point sources distribution to an
$\alpha$-stable model. The method is tested in the presence of a
Gaussian background. These approaches offer interesting possibilities
for the extraction of statistical information from an unresolved
background of point sources. However, the study of their performance
on more realistic conditions -- that take into account, for instance,
the presence of anisotropic noise and non-Gaussian foregrounds --
remains to be done.

Another interesting possibility is the study of the bispectrum of
undetected point sources since this quantity will depend on the
characteristics of the underlying population of extragalactic point
sources ~\cite{bb:kom01,bb:arg03}. It also
provides with an estimation of the level of the contamination
introduced by the residual point sources. In particular, \cite{bb:kom03}
estimated the power spectrum of residual point sources in
the WMAP data through the measurement of the bispectrum.

One may also use statistical information of the unresolved sources to
identify which type of emission is present. In particular,
\cite{bb:rub03} presents a detailed study of the contribution of the
thermal SZ emission and of the extragalactic point sources to the
probability distribution of the brightness map. The different
imprints left by these two emissions would allow one to discriminate
whether the excess of power found at small scales in some CMB data is
due to an unresolved background of point sources or to the presence of
unresolved SZ clusters. An alternative study based on the analysis of the
Gaussianity of the wavelet coefficients has also been carried out to
explain this excess of power~\cite{bb:die04}.

Regarding the kinetic SZ effect, we have already mentioned that the
detection of peculiar velocities of individual clusters will be a very
challenging task. However, one could infer bulk flows on large scales
what would provide valuable cosmological information. This point has
been addressed by different authors, including
\cite{bb:kas00,bb:agh01,bb:atr04}.

\section{Conclusions}
\label{sec:conclusions}

The development of techniques for the extraction of compact sources in
CMB observations has become a very relevant and active topic. This is
due to the necessity of cleaning the CMB maps from astrophysical
contaminants that would impair our ability to extract all the valuable
information encoded in the cosmological signal but also because the
recovered catalogues of point sources and/or SZ clusters would contain
themselves extremely relevant astrophysical and cosmological
information.

An important effort has been done in the last years towards the
development of more powerful and sophisticated tools to extract compact
sources. Many of them have been tested on simulated Planck
observations, showing their potentiality. However some important work
still remains to be done. First of all, in some cases, quite ideal
conditions have been assumed. For instance, it is commonly assumed
that the cluster profile is known but, in general, this will not be
the case for real data. Other methods have been applied to simulations
that do not include foreground emissions. Therefore, these and other
problematics -- beam asymmetry, extension to the sphere, relativistic
effects, anisotropic noise, etc. -- present on real data should be
taken into account to establish the true performance of the developed
methods. Also, the methods do not always use all the available
information present in the data. For instance, if multifrequency
observations are available, it would be useful to include this
multifrequency information in the detection of extragalactic point
sources even if they do not follow a simple well-known spectral
law. The final and most important step would be to apply these
techniques to real CMB data (e.g. WMAP) as they become available.

We would also like to point out that many polarisation experiments are
currently planned (or already operating) which will provide with a
wealth of information about our universe \cite{bb:cha}. Given the
weakness of the cosmological signal in polarisation and the current
lack of knowledge regarding the foreground emissions, a careful
process of cleaning of the CMB polarisation maps is even more critical
than for the intensity case. However no techniques have been yet
specifically developed to extract compact sources from polarisation
CMB observations. Therefore it is crucial to extend some of the
current methods -- or to develop new ones -- to deal with this type of
maps.

Finally, a very critical issue is to assess which is the impact of
possible residuals left in the CMB data after applying these
techniques~\cite{bb:bar05}. In particular, it is very important
to control the effect of undetected sources, or even possible
artifacts introduced in the image after subtracting the signals, on
the estimation of the power spectrum of the CMB. In addition, this
process should not modify the underlying CMB temperature distribution,
since it would impair our ability to perform Gaussianity analyses of the
CMB (or even lead us to wrong conclusions), which are of great
importance to learn about the structure formation of our universe.

\section*{Acknowledgements}
RBB acknowledges the Universidad de Cantabria and the Ministerio de
Educaci\'on y Ciencia for a Ram\'on y Cajal contract and J.M. Diego,
D. Herranz, M. L\'opez-Caniego, E. Mart\'{\i}nez-Gonz\'alez, J.L. Sanz
and P. Vielva for their help in the ellaboration of these lecture
notes. I would also like to thank the organisers for inviting me to
participate in a very fruitful and interesting school.

%
%
%

%

\begin{thebibliography}{99.}
%
%
%

\bibitem{bb:agh01} N. Aghanim, K. M. G\'orski, J.-L. Puget: A\&A,
\textbf{374}, 1 (2001)

\bibitem{bb:arg03} F. Arg\"ueso, J. Gonz\'alez-Nuevo, L. Toffolatti:
ApJ, \textbf{598}, 86 (2003)

\bibitem{bb:atr04} F. Atrio-Barandela, A. Kashlinsky, J. P. M\"ucket:
ApJ, \textbf{601}, 111 (2004)

\bibitem{bb:bac00} C. Baccigalupi, L. Bedini, C. Burigana, G. De
Zotti, A. Farusi, D. Maino, M. Maris, F. Perrota, E. Salerno,
L. Toffolatti, A. Tonazzini: MNRAS, \textbf{318}, 769 (2000)

\bibitem{bb:ban03} A. J. Banday, C. Dickinson, R. D. Davies,
R. J. Davis, K. M. G\'orski: MNRAS, \textbf{345}, 897 (2003)

\bibitem{bb:barn04} V. E. Barnard, P. Vielva, D. P. I. Pierce-Price,
A. W. Blain, R. B. Barreiro, J. S. Richer, C. Qualtrough: MNRAS,
\textbf{352}, 961 (2004)

\bibitem{bb:bar03} R. B. Barreiro, J. L. Sanz, D. Herranz,
E. Mart{\'\i}nez-Gonz\'alez: MNRAS, \textbf{342}, 119 (2003)

\bibitem{bb:bar04} R. B. Barreiro, M. P. Hobson, A. J. Banday,
A. N. Lasenby, V. Stolyarov, P. Vielva, K. M. G\'orski: MNRAS,
\textbf{351}, 515 (2004)

\bibitem{bb:bar05} R. B. Barreiro, E. Mart{\'\i}nez-Gonz\'alez,
P. Vielva, M. P. Hobson: MNRAS, submitted, preprint (astro-ph/0503039)

\bibitem{bb:bed05} L. Bedini, D. Herranz, E. Salerno, C. Baccigalupi,
E. E. Kuruoglu, A. Tonazzini: EURASIP Journal on Applied Signal
Processing (Special Issue on Applications of Signal Processing in
Astrophysics and Cosmology), \textbf{15}, 2400 (2005)

\bibitem{bb:ben03} C. L. Bennett et al.: ApJS, \textbf{148}, 97 (2003)

\bibitem{bb:ber96} E. Bertin, S. Arnouts: A\&AS, \textbf{117}, 393
(1996) 

\bibitem{bb:bir99} M. Birkinshaw: PhR, \textbf{310}, 97 (1999)

\bibitem{bb:bou96} F. R. Bouchet, R. Gispert, J. L. Puget: The
MM/SUB-MM foregrounds and future CMB space missions. In: \textit{ The
COBE workshop: Unveiling the cosmic infrared background} AIP
Conference Proceedings, vol 348, ed by E. Dwek (1996), pp 255 -- 270

\bibitem{bb:bou99} F. R. Bouchet, R. Gispert: New Astr., \textbf{4},
443 (1999) 

\bibitem{bb:bur98} C. S. Burrus, R. A. Gopinath, H. Guo:
\textit{Introduction to Wavelets and Wavelet Transforms. A primer}
(Prentince-Hall, Upper Saddle River, New Jersey 1998)

\bibitem{bb:car02} J. E. Carlstrom, G. P. Holder,
E. D. Reese: AR\&A\&A, \textbf{40}, 643 (2002)

\bibitem{bb:cay00} L. Cay\'on, J.L. Sanz, R.B. Barreiro, E. Mart\'\i
nez-Gonz\'alez, P. Vielva, L. Toffolatti, J. Silk, J.M. Diego,
F. Arg\"ueso: MNRAS, \textbf{315}, 757 (2000)

\bibitem{bb:cha} A. Challinor: \textit{Cosmic microwave background
polarization analysis}, in this volume

\bibitem{bb:chi02} L.-Y. Chiang, H. E. Jorgensen, I. P. Naselsky,
P. D. Naselsky, I. D. Novikov, P. R. Christensen: MNRAS, \textbf{335},
1054 (2002)

\bibitem{bb:cob03} K. Coble et al., ApJS, submitted, preprint
(astro-ph/0301599) 

\bibitem{bb:dam97a} F. Damiani, A. Maggio, G. Micela, S. Sciortino:
ApJ, \textbf{483}, 350 (1997)

\bibitem{bb:dam97b} F. Damiani, A. Maggio, G. Micela, S. Sciortino:
ApJ, \textbf{483}, 370 (1997)


\bibitem{bb:dau92} I. Daubechies: \textit{Ten Lectures on Wavelets} (S.I.A.M.,
Philadelphia, 1992)

\bibitem{bb:del03} J. Dellabrouille, J. F. Cardoso, G. Patanchon:
MNRAS, \textbf{346}, 1089 (2003)

\bibitem{bb:deo02} A. de Oliveira-Costa et al.: ApJ, \textbf{567}, 363
(2002) 

\bibitem{bb:deo04} A. de Oliveira-Costa et al., M. Tegmark,
R. D. Davies, C. M. Guti\'errez, A. N. Lasenby, R. Rebolo,
R. A. Watson: ApJ, \textbf{606}, L89 (2004)

\bibitem{bb:dez99} G. de Zotti, L. Toffolatti, F. Arg\"ueso,
R. D. Davies, P. Mazzotta, R. B. Partridge, G. F. Smoot, N. Vittorio:
The Planck Surveyor Mission: Astrophysical Prospects. In \textit{3K
Cosmology, Proceedings of the EC-TMR Conference held in Rome, Italy,
October, 1998}, American Institute of Physics, vol. 476, ed by
L. Maiani, F. Melchiorri, N. Vittorio (Woodbury, N.Y. 1999) pp 204 --
223

\bibitem{bb:dez05} G. de Zotti, R. Ricci, D. Mesa, L. Silva,
P. Mazzotta, L. Toffolatti, J. Gonz\'alez-Nuevo: A\&A, \textbf{431},
893 (2005) 

\bibitem{bb:die02} J. M. Diego, P. Vielva,
E. Mart{\'\i}nez-Gonz\'alez, J. Silk, J.L. Sanz: MNRAS, \textbf{336},
1351 (2002)

\bibitem{bb:die04} J. M. Diego, P. Vielva, E. Mar{\'\i}nez-Gonz\'alez,
J. Silk: preprint (astro-ph/0403561)

\bibitem{bb:dol03} H. Dole, G. Lagache, J.-L. Puget: ApJ,
\textbf{585}, 617 (2003)

\bibitem{bb:dra98} B. T. Draine, A. Lazarian: ApJ, \textbf{494}, L19
(1998)

\bibitem{bb:dra98b} B. T. Draine, A. Lazarian: ApJ, \textbf{508}, 157 (1998)

\bibitem{bb:eri04} H. K. Eriksen, A. J. Banday, K. M. G\'orski,
P. B. Lilje: ApJ, \textbf{612}, 633 (2004)

\bibitem{bb:eri05} H. K. Eriksen et al.: ApJ, submitted (2005),
preprint (astro-ph/0508268)

\bibitem{bb:fin04} D. P. Finkbeiner, G. I. Langston, A. H. Minter:
ApJ, \textbf{617}, 350 (2004)

\bibitem{bb:for04} O. Forni, N. Aghanim: A\&A, \textbf{420}, 49 (2004)

\bibitem{bb:gon05}  J. Gonz\'alez-Nuevo, L. Toffolatti, F. Arg\"ueso:
ApJ, \textbf{621}, 1 (2005)

\bibitem{bb:gra04} G. L. Granato, G. de Zotti, L. Silva, A. Bressan,
L. Danese: ApJ, \textbf{600}, 580 (2004)

\bibitem{bb:gui99} B. Guiderdoni: Far-Infrared Point Sources. In:
\textit{Microwave Foregrounds}, ASP Conference Series, vol 181, ed by
A. de Oliveira-Costa, M. Tegmark (1999), pp 173 -- 198

\bibitem{bb:hae96} M. G. Haehnelt \& M Tegmark: MNRAS, \textbf{279},
545 (1996)

\bibitem{bb:her02a} D. Herranz, J.L. Sanz, M.P. Hobson, R.B. Barreiro,
J.M. Diego, E. Mart\'\i nez-Gonz\'alez, A.N. Lasenby: MNRAS,
\textbf{336}, 1057 (2002)

\bibitem{bb:her02b} D. Herranz, J. Gallegos, J.L. Sanz,
E. Mart{\'\i}nez-Gonz\'alez: MNRAS, \textbf{334}, 353 (2002)

\bibitem{bb:her02c} D. Herranz, J.L. Sanz, R.B. Barreiro,
E. Mart{\'\i}nez-Gonz\'alez: ApJ, \textbf{580}, 610 (2002)

\bibitem{bb:her02d} D. Herranz: Analysis of the anisotropies in the
Cosmic Microwave Background Radiation using Adaptive Filters.
PhD Thesis, Universidad de Cantabria, Spain (2002)

\bibitem{bb:her04} D. Herranz, E. E. Kuruoglu, L. Toffolatti: A\&A,
\textbf{424}, 1081 (2004)

\bibitem{bb:her05} D. Herranz, J.L. Sanz, R.B. Barreiro,
M. L\'opez-Caniego: MNRAS, \textbf{356}, 944 (2005)

\bibitem{bb:hob98} M. P. Hobson, A. W. Jones, A. N. Lasenby,
F. R. Bouchet: MNRAS, \textbf{300}, 1 (1998)

\bibitem{bb:hob99} M. P. Hobson, R. B. Barreiro, L. Toffolatti,
A. N. Lasenby, J. L. Sanz, A. W. Jones, F. R. Bouchet: MNRAS,
\textbf{306}, 232 (1999)

\bibitem{bb:hob03} M.P. Hobson, C. McLachlan: MNRAS, \textbf{338}, 765
(2003)

\bibitem{bb:kas00} A. Kashlinsky, F. Atrio-Barandela: ApJ,
\textbf{536}, 67 (2000)

\bibitem{bb:knu05} K. K. Knudsen, V. E. Barnard, R. P. van der Werf,
P. Vielva, J.-P. Kneib, A. W. Blain, R. B. Barreiro, R. J. Ivison,
I. R. Smail, J. A. Peacock: MNRAS, submitted (2005)

\bibitem{bb:kom01} E. Komatsu, D. N. Spergel: PhRvD, \textbf{63},
063002 (2001)

\bibitem{bb:kom03} E. Komatsu et al.: ApJS, \textbf{148}, 119 (2003)

\bibitem{bb:lag03} G. Lagache: A\&A, \textbf{405}, 813 (2003)

\bibitem{bb:lop04} M. L\'opez-Caniego, D. Herranz, R. B. Barreiro,
J. L. Sanz: SPIE, \textbf{5299}, 145 (2004)

\bibitem{bb:lop05} M. L\'opez-Caniego, D. Herranz, R. B. Barreiro,
J. L. Sanz: MNRAS, \textbf{359}, 993 (2005)

\bibitem{bb:lop05b} M. L\'opez-Caniego, D. Herranz, J. L. Sanz,
R. B. Barreiro: EURASIP Journal on Applied Signal Processing (Special
Issue on Applications of Signal Processing in Astrophysics and
Cosmology), \textbf{15}, 1 (2005)

\bibitem{bb:lop05c} M. L\'opez-Caniego, J. L. Sanz, D. Herranz,
J. Gonz\'alez-Nuevo, R. B. Barreiro, E. E. Kuruoglu:
\textit{Non-linear fusion of images and the detection of point
sources}. In: IEEE proceedings of the International Workshop on
Nonlinear Signal and Image Processing, in press (2005)

\bibitem{bb:lop05d} M. L\'opez-Caniego, J. L. Sanz, D. Herranz,
R. B. Barreiro, J. Gonz\'alez-Nuevo: \textit{Linear and quadratic
fusion of images: detection of point sources}. In: Proceedings of the
13th European Signal Processing Conference (EUSIPCO 2005), in press
(2005) 

\bibitem{bb:mai02} D. Maino, A. Farusi, C. Baccigalupi, F. Perrotta,
A. J. Banday, L. Bedini, C. Burigana, G. de Zotti, K. M. G\'orski,
E. Salerno: MNRAS, \textbf{334}, 53 (2002)

\bibitem{bb:mai03} D. Maino, A. J. Banday, C. Baccigalupi,
F. Perrotta, K. M. G\'orski: MNRAS, \textbf{344}, 544 (2003)

\bibitem{bb:mar02} E. Mart{\'\i}nez-Gonz\'alez, J. E. Gallegos,
F. Arg\"ueso, L. Cay\'on, J. L. Sanz: MNRAS, \textbf{336}, 22 (2002) 

\bibitem{bb:mar03} E. Mart{\'\i}nez-Gonz\'alez, J. M. Diego,
P. Vielva, J. Silk: MNRAS, \textbf{345}, 1101 (2003)

\bibitem{bb:muk02} P. Mukherjee, B. Dennison, B. Ratra,
J. H. Simonetti, K. Ganga, J.-C. Hamilton: ApJ, \textbf{579}, 83
(2002)

\bibitem{bb:nas02} P. Naselsky, D. Novikov, J. Silk: MNRAS, \textbf{335}, 550
(2002)

\bibitem{bb:pat05} G. Patanchon, J. F. Cardoso, J. Delabrouille,
P. Vielva: MNRAS, in press (2005), preprint (astro-ph/0410280) 

\bibitem{bb:planck} Planck Mission: http://www.rssd.esa.int/Planck

\bibitem{bb:pie03} E. Pierpaoli: ApJ, \textbf{589}, 58 (2003)

\bibitem{bb:pie05} E. Pierpaoli, S. Anthoine, K. Huffenberger,
I. Daubechies: MNRAS, \textbf{359}, 261 (2005)


\bibitem{bb:rub03} J. A. Rubi\~no-Mart\'{\i}n, R. A. Sunyaev: MNRAS,
\textbf{344}, 1155 (2003)

\bibitem{bb:san01} J. L. Sanz, D. Herranz \& E. Mart\'{\i}nez-Gonz\'alez:
ApJ, \textbf{552}, 484 (2001)

\bibitem{bb:sch04}  B. M. Schaefer, C. Pfrommer, R. Hell,
M. Bartelmann: MNRAS, submitted, preprint (astro-ph/0407090)

\bibitem{bb:sch03} A. E. Schulz \& M. White: ApJ, \textbf{586}, 723 (2003)

\bibitem{bb:sto02} V. Stolyarov, M. P. Hobson, M. A. J. Ashdown,
A. N. Lasenby: MNRAS, \textbf{336}, 97 (2002)

\bibitem{bb:sto05} V. Stolyarov, M. P. Hobson, A. N. Lasenby,
R. B. Barreiro: MNRAS, \textbf{357}, 145 (2005)

\bibitem{bb:sz70} R. A. Sunyaev, Ya. B. Zel'dovich: Ap\&SS,
\textbf{7}, 3 (1970)

\bibitem{bb:sz72} R. A. Sunyaev, Ya. B. Zel'dovich: CoASP, \textbf{4},
173 (1972)

\bibitem{bb:sz80} R. A. Sunyaev, Ya. B. Zel'dovich: MNRAS, \textbf{190},
413 (1980)

\bibitem{bb:teg96} M. Tegmark \& G. Efstathiou: MNRAS, \textbf{281},
1297 (1996)

\bibitem{bb:teg98} M. Tegmark, A. de Oliveira-Costa: ApJ,
\textbf{500}, L83 (1998)

\bibitem{bb:teg99} M. Tegmark, D. J. Eisenstein, W. Hu, A. de
Oliveira-Costa: Overview of the foregrounds and their impact. In:
\textit{Microwave Foregrounds}, ASP Conference Series, vol 181, ed by
A. de Oliveira-Costa, M. Tegmark (1999), pp 3 -- 58

\bibitem{bb:teg03} M. Tegmark, A. de Oliveira-Costa,
A. J. S. Hamilton: Phys. Rev. D., \textbf{68}, 123523 (2003)

\bibitem{bb:tof98} L. Toffolatti, F. Arg\"ueso-G\'omez, G. De Zotti,
P. Mazzei, A. Francheschini, L. Danese, C. Burigana: MNRAS,
\textbf{297}, 117 (1998)

\bibitem{bb:tuc04} M. Tucci, E. Mart\'\i nez-Gonz\'alez, L. Toffolatti,
J. Gonz\'alez-Nuevo, G. De Zotti: MNRAS, \textbf{349}, 1267 (2004)

\bibitem{bb:vie01a} P. Vielva, E. Mart{\'\i}nez-Gonz{\'a}lez, L. Cay\'on,
J.L. Sanz, L. Toffolatti: MNRAS, \textbf{326}, 181 (2001)

\bibitem{bb:vie01b} P. Vielva, R.B. Barreiro, M.P. Hobson,
E. Mart{\'\i}nez-Gonz{\'a}lez, A.N. Lasenby, J.L. Sanz, L. Toffolatti:
MNRAS, \textbf{328}, 1 (2001)
%
\bibitem{bb:vie03} P. Vielva, E. Mart{\'\i}nez-Gonz{\'a}lez,
J.E. Gallegos, L. Toffolatti L., J.L. Sanz: MNRAS, \textbf{344}, 89
(2003) 





\end{thebibliography}
%



\printindex
\end{document}